\documentclass[preprint]{aastex}

\begin{document}

\title{TY Pup: a low-mass-ratio and deep contact binary as a progenitor candidate of luminous red novae}

\author{T. Sarotsakulchai\altaffilmark{1,4,5},
S.-B. Qian\altaffilmark{1,2,3,4},
B. Soonthornthum\altaffilmark{5},
X. Zhou\altaffilmark{1,2,3},
J. Zhang\altaffilmark{1,2,3}, \\
D. E. Reichart\altaffilmark{6},
J. B. Haislip\altaffilmark{6},
V. V. Kouprianov\altaffilmark{6} and
S. Poshyachinda\altaffilmark{5}}

\altaffiltext{1}{Yunnan Observatories, Chinese Academy of Sciences, 650216 Kunming, China} \email{huangbinghe@ynao.ac.cn}
\altaffiltext{2}{Key Laboratory of the Structure and Evolution of Celestial Objects, Chinese Academy of Sciences, 650216 Kunming, China}
\altaffiltext{3}{Center for Astronomical Mega-Science, Chinese Academy of Sciences, 20A Datun Rd., Chaoyang District, Beijing, 100012, China}
\altaffiltext{4}{University of Chinese Academy of Sciences, 19 A Yuquan Rd., Shijingshan, 100049 Beijing, China}
\altaffiltext{5}{National Astronomical Research Institute of Thailand, Ministry of Science and Technology, Bangkok, Thailand}
\altaffiltext{6}{Department of Physics and Astronomy, University of North Carolina, CB \#3255, Chapel Hill, NC 27599, USA}


\begin{abstract}
TY Pup is a well-known bright eclipsing binary with an orbital period of 0.8192 days. New light curves in $B, V, (RI)_C$ bands were obtained with the 0.61-m reflector robotic telescope (PROMPT-8) at CTIO in Chile during 2015 and 2017. By analysing those photometric data with the W-D method, it is found that TY Pup is a low-mass-ratio ($q \sim$ 0.184) and deep contact binary with a high fill-out factor ($84.3\,\%$). An investigation of all available times of minimum light including three new ones obtained with the 60-cm and the 1.0-m telescopes at Yunnan Observatories in China indicates that the period change of TY Pup is complex. An upward parabolic variation in the $O-C$ diagram is detected to be superimposed on a cyclic oscillation. The upward parabolic change reveals a long-term continuous increase in the orbital period at a rate of $\mathrm{d}P/\mathrm{d}t = 5.57(\pm 0.08)\times10^{-8}$ d $\textrm{yr}^{-1}$. The period increase can be explained by mass transfer from the less massive component ($M_2 \sim 0.3 M_{\odot}$) to the more massive one ($M_1 \sim 1.65 M_{\odot}$). The binary will be merging when it meets the criterion that the orbital angular momentum is less than 3 times the total spin angular momentum, i.e., $J_{orb} < 3J_{rot}$. This suggests that the system will finally merge into a rapid-rotating single star and may produce a luminous red nova. The cyclic oscillation in the $O-C$ diagram can be interpreted by the light-travel time effect (LITE) via the presence of a third body.
\end{abstract}

\section{Introduction}

W UMa-type stars are short-period ($P<1$\,day) binaries where both component stars are filling the critical Roche lobe and possess a common envelope (e.g., Qian et al. 2017). They evolved from detached binary stars via angular momentum loss and/or a case A mass transfer (e.g., Qian et al. 2018). Low-mass-ratio and deep contact binaries are on the late evolutionary state of contact binary systems. They have a high fill-out factor ($f>50\%$) and a very low mass ratio ($q<0.25$) (Qian et al. 2005). This type of binary may be the progenitors of single rapidly-rotating stars (e.g. Kandulapati et al., 2015; Sriram et al., 2016, 2017; Li et al. 2017; Liao et al. 2017; Samec et al. 2011, 2018) and will produce a new type of stellar outburst, i.e., luminous red novae (e.g., Tylenda et al. (2011); Stepien (2011); Zhu et al. 2016; Molnar et al. (2017)). Contact binary V1309 Sco is an example of progenitors for this eruption. These properties make them an important source to understand the merging of binaries and to investigate the structure and evolution of contact binaries at the late stage. On the other hand, W UMa-type binaries have the shortest orbital period and the lowest angular momentum among main-sequence binary stars. Searching for and studying the third components of such systems can also provide more information of their formation and evolution because they may have played an important role during the origin and evolution of contact binaries by removing angular momentum from the central binaries (Qian et al. 2013a).

TY Pup (HIP 36683, HD 60265) is one of the bright contact binaries in southern hemisphere, which was discovered by Hertzsprung (1928). Campbell (1928) made the first photometric measurements and derived its period as 0.58071564\,days. The first spectroscopic observations were performed by Struve (1950) and found two periodicities with periods of 0.58 and 9.7\,days, respectively. Later, new photometric observations were carried out by Huruhata et. al. (1957), but the results were unable to confirm either Struve's value of zero epoch or his secondary period of 9.7 days. The correct period of TY Pup was derived by Van Houten (1971) as 0.819235 days, which fitted well for both photometric observations by Huruhata et al. (1957) and the spectroscopic one by Struve (1950). Struve (1950) classified its spectral type as A9n, but Duerbeck \& Rucinski (2007) reported that the spectral classification of HDH (Michigan Catalogue of HD Stars) is F3V and agrees with the Tycho-2, $B-V$=0.39 (Hog et al. 2000). The radial-velocity studies by Duerbeck \& Rucinski (2007) suggested that TY Pup is a typical A-subtype contact binary with a mass ratio of $q$ = 0.25.

\begin{table}
\scriptsize
\caption{Coordinates of TY Pup, the comparison, and the check stars.}
\begin{center}
\begin{tabular}{llccrcc}\hline\hline
Targets & name&$\alpha_{2000}$ & $\delta_{2000}$ & $mag (V)$& $B-V$& $J-H$
\\\hline
Binary star &TY Pup & $07^{\textrm{h}}32^{\textrm{m}}46^{\textrm{s}}.3$ & $-20^\circ47^{\prime}29^{\prime\prime}.5$ &8.62 &0.39 &0.169\\
The comparison & HD 60342& $07^{\textrm{h}}33^{\textrm{m}}10^{\textrm{s}}.2$ & $-20^\circ42^{\prime}13^{\prime\prime}.1$ &8.56 &-0.06 &-0.067\\
The check &TYC 5991-1892-1 & $07^{\textrm{h}}32^{\textrm{m}}37^{\textrm{s}}.9$ & $-20^\circ45^{\prime}05^{\prime\prime}.1$ &10.21 &0.35 &0.188\\
\hline
\end{tabular}
\end{center}
\end{table}

Recently, based on V-band observations obtained by the All Sky Automated Survey (ASAS,  Pojmanski 1997, 2002), the physical parameters of TY Pup were determined by Deb \& Singh (2011). Variations in light curves were found, but did not show O'Connell effect (O'Connell 1951). For the orbital period study, times of minimum light of TY Pup were collected and investigated by several authors (e.g. Gu et al. (1993); Berdnikov \& Turner (1995)), they gave a linear ephemeris with no changes in the orbital period. However, Qian (2001) found that the period was increasing continuously at a rate of $\mathrm{d}P/\mathrm{d}t = 1.66\times10^{-7}$ d $\textrm{yr}^{-1}$. In this paper, we present new CCD observations and their photometric solutions. Then the changes in the orbital period are investigated based on all available eclipse times which shows a combination of a cyclic variation and a continuously increasing period. We detect that TY Pup is a low-mass-ratio and deep contact binary with an additional companion and it may be a progenitor candidate of luminous red novae.

\section{New CCD photometric Observations}

The first set of light curves of TY Pup in $BV(RI)_C$ bands were carried out for several nights from January to February 2015 with the back illuminated Apogee F42 2048$\times$2048 CCD attached to the 0.6 m Cassegrain reflecting telescope of PROMPT-8\footnote{PROMPT-8 is the Thai Southern Hemisphere Telescope (TST), operated in collaboration between National Astronomical Research Institute of Thailand (NARIT) and the University of North Carolina (UNC) at Chapel Hill in a part of the UNC-led PROMPT project, http://skynet.unc.edu.} robotic telescope. The telescope is located at the Cerro Tololo Inter-American Observatory (CTIO) in Chile. The web-based SKYNET client allowed us to request and retrieve image remotely via the internet. SKYNET system also provided nightly calibration images, including bias, dark, and flat-field images (Layden et al. 2010). All CCD reductions and aperture photometry measurements were done with standard procedure packages of IRAF\footnote{The Image Reduction and Analysis Facility (IRAF), http://iraf.noao.edu.}.

\begin{table}
\caption{CCD observations in BVRI bands for TY Pup observed in 2015}
\begin{center}
\begin{tiny}
\begin{tabular}{cccccccccccccc}\hline\hline
HJD&$\Delta B$ &HJD &$\Delta V$ &HJD &$\Delta R$ &HJD&$\Delta I$&HJD&$\Delta B$ &HJD &$\Delta V$ &HJD &$\Delta R$ \\
-2457000&mag&&&&&&&&&&&&\\
\hline
39.6317	&	0.583	&	39.6321	&	0.128	&	39.6325	&	-0.139	&	39.6329	&	-0.386	&	39.7541	&	0.220	&	39.7545	&	-0.206	&	39.7549	&	-0.482	\\
39.6333	&	0.577	&	39.6337	&	0.104	&	39.6341	&	-0.152	&	39.6345	&	-0.406	&	39.7558	&	0.215	&	39.7562	&	-0.204	&	39.7566	&	-0.477	\\
39.6349	&	0.571	&	39.6353	&	0.107	&	39.6358	&	-0.157	&	39.6362	&	-0.416	&	39.7574	&	0.224	&	39.7578	&	-0.208	&	39.7582	&	-0.465	\\
39.6367	&	0.553	&	39.6371	&	0.103	&	39.6375	&	-0.165	&	39.6379	&	-0.427	&	39.7591	&	0.206	&	39.7595	&	-0.213	&	39.7599	&	-0.480	\\
39.6384	&	0.548	&	39.6388	&	0.074	&	39.6392	&	-0.176	&	39.6396	&	-0.434	&	39.7608	&	0.207	&	39.7612	&	-0.213	&	39.7616	&	-0.494	\\
39.6401	&	0.551	&	39.6405	&	0.078	&	39.6408	&	-0.189	&	39.6413	&	-0.438	&	39.7624	&	0.228	&	39.7627	&	-0.221	&	39.7632	&	-0.483	\\
39.6417	&	0.536	&	39.6422	&	0.074	&	39.6426	&	-0.196	&	39.6430	&	-0.423	&	39.7641	&	0.211	&	39.7645	&	-0.222	&	39.7649	&	-0.497	\\
39.6434	&	0.520	&	39.6439	&	0.062	&	39.6443	&	-0.200	&	39.6447	&	-0.443	&	39.7658	&	0.210	&	39.7662	&	-0.198	&	39.7666	&	-0.492	\\
39.6452	&	0.516	&	39.6456	&	0.050	&	39.6459	&	-0.203	&	39.6463	&	-0.449	&	39.7716	&	0.189	&	39.7720	&	-0.232	&	39.7724	&	-0.500	\\
39.6468	&	0.505	&	39.6472	&	0.062	&	39.6476	&	-0.225	&	39.6481	&	-0.452	&	39.7733	&	0.200	&	39.7737	&	-0.207	&	39.7741	&	-0.482	\\
39.6485	&	0.505	&	39.6489	&	0.047	&	39.6494	&	-0.214	&	39.6498	&	-0.485	&	39.7750	&	0.192	&	39.7754	&	-0.210	&	39.7758	&	-0.488	\\
39.6502	&	0.486	&	39.6506	&	0.041	&	39.6509	&	-0.217	&	39.6514	&	-0.479	&	39.7767	&	0.193	&	39.7771	&	-0.209	&	39.7775	&	-0.494	\\
39.6518	&	0.491	&	39.6522	&	0.036	&	39.6527	&	-0.243	&	39.6531	&	-0.486	&	39.7784	&	0.191	&	39.7788	&	-0.205	&	39.7792	&	-0.502	\\
39.6535	&	0.478	&	39.6540	&	0.030	&	39.6544	&	-0.219	&	39.6548	&	-0.478	&	39.7801	&	0.192	&	39.7805	&	-0.220	&	39.7809	&	-0.507	\\
39.6553	&	0.472	&	39.6557	&	-0.001	&	39.6561	&	-0.242	&	39.6565	&	-0.484	&	39.7818	&	0.195	&	39.7822	&	-0.218	&	39.7826	&	-0.498	\\
39.6569	&	0.469	&	39.6573	&	0.008	&	39.6577	&	-0.238	&	39.6581	&	-0.493	&	39.7835	&	0.200	&	39.7839	&	-0.232	&	39.7843	&	-0.493	\\
39.6586	&	0.459	&	39.6590	&	-0.014	&	39.6594	&	-0.256	&	39.6597	&	-0.515	&	39.7852	&	0.172	&	39.7855	&	-0.248	&	39.7860	&	-0.502	\\
39.6602	&	0.462	&	39.6606	&	-0.010	&	39.6610	&	-0.271	&	39.6614	&	-0.509	&	39.7868	&	0.187	&	39.7873	&	-0.209	&	39.7877	&	-0.492	\\
39.6619	&	0.448	&	39.6623	&	-0.013	&	39.6627	&	-0.269	&	39.6630	&	-0.524	&	39.7885	&	0.158	&	39.7889	&	-0.210	&	39.7893	&	-0.474	\\
39.6635	&	0.445	&	39.6639	&	-0.021	&	39.6643	&	-0.287	&	39.6647	&	-0.534	&	39.7902	&	0.190	&	39.7907	&	-0.218	&	39.7912	&	-0.491	\\
39.6651	&	0.423	&	39.6655	&	-0.031	&	39.6659	&	-0.283	&	39.6665	&	-0.509	&	39.7922	&	0.181	&	39.7926	&	-0.222	&	39.7930	&	-0.505	\\
39.6683	&	0.417	&	39.6692	&	-0.036	&	39.6722	&	-0.307	&	39.6727	&	-0.552	&	39.7939	&	0.182	&	39.7944	&	-0.208	&	39.7948	&	-0.479	\\
39.6731	&	0.418	&	39.6735	&	-0.064	&	39.6739	&	-0.320	&	39.6744	&	-0.558	&	39.7957	&	0.166	&	39.7961	&	-0.240	&	39.7965	&	-0.499	\\
39.6749	&	0.400	&	39.6753	&	-0.070	&	39.6758	&	-0.330	&	39.6762	&	-0.555	&	39.7975	&	0.166	&	39.7978	&	-0.221	&	39.7983	&	-0.490	\\
39.6767	&	0.384	&	39.6771	&	-0.054	&	39.6775	&	-0.311	&	39.6779	&	-0.588	&	39.7992	&	0.212	&	39.7996	&	-0.218	&	39.8001	&	-0.478	\\
39.6784	&	0.391	&	39.6789	&	-0.068	&	39.6792	&	-0.319	&	39.6797	&	-0.558	&	39.8010	&	0.204	&	39.8014	&	-0.198	&	39.8019	&	-0.516	\\
39.6801	&	0.377	&	39.6806	&	-0.074	&	39.6809	&	-0.335	&	39.6813	&	-0.574	&	39.8028	&	0.209	&	39.8033	&	-0.224	&	39.8037	&	-0.500	\\
39.6818	&	0.374	&	39.6822	&	-0.083	&	39.6827	&	-0.339	&	39.6831	&	-0.579	&	39.8227	&	0.245	&	39.8232	&	-0.188	&	39.8236	&	-0.468	\\
39.6836	&	0.365	&	39.6911	&	-0.111	&	39.6972	&	-0.383	&	39.6977	&	-0.626	&	39.8245	&	0.238	&	39.8250	&	-0.192	&	39.8253	&	-0.452	\\
39.6982	&	0.320	&	39.6986	&	-0.119	&	39.6990	&	-0.383	&	39.6995	&	-0.634	&	39.8262	&	0.243	&	39.8267	&	-0.169	&	39.8270	&	-0.455	\\
39.6999	&	0.301	&	39.7004	&	-0.113	&	39.7010	&	-0.383	&	39.7013	&	-0.623	&	39.8279	&	0.248	&	39.8283	&	-0.184	&	39.8287	&	-0.478	\\
39.7019	&	0.303	&	39.7024	&	-0.120	&	39.7028	&	-0.394	&	39.7032	&	-0.640	&	39.8296	&	0.230	&	39.8300	&	-0.152	&	39.8305	&	-0.466	\\
39.7037	&	0.292	&	39.7042	&	-0.119	&	39.7046	&	-0.395	&	39.7050	&	-0.643	&	39.8314	&	0.260	&	39.8318	&	-0.189	&	39.8322	&	-0.484	\\
39.7055	&	0.296	&	39.7060	&	-0.122	&	39.7065	&	-0.406	&	39.7069	&	-0.655	&	39.8331	&	0.233	&	39.8335	&	-0.174	&	39.8339	&	-0.451	\\
39.7074	&	0.293	&	39.7078	&	-0.136	&	39.7082	&	-0.396	&	39.7087	&	-0.648	&	39.8349	&	0.250	&	39.8353	&	-0.152	&	39.8357	&	-0.461	\\
39.7092	&	0.287	&	39.7096	&	-0.130	&	39.7100	&	-0.408	&	39.7104	&	-0.630	&	39.8366	&	0.243	&	39.8370	&	-0.181	&	39.8375	&	-0.445	\\
39.7110	&	0.278	&	39.7115	&	-0.136	&	39.7120	&	-0.412	&	39.7124	&	-0.657	&	39.8384	&	0.269	&	39.8388	&	-0.191	&	39.8391	&	-0.455	\\
39.7129	&	0.286	&	39.7133	&	-0.137	&	39.7137	&	-0.417	&	39.7141	&	-0.646	&	39.8400	&	0.254	&	39.8404	&	-0.153	&	39.8408	&	-0.443	\\
39.7146	&	0.279	&	39.7150	&	-0.149	&	39.7153	&	-0.424	&	39.7157	&	-0.665	&	39.8417	&	0.276	&	39.8421	&	-0.178	&	39.8426	&	-0.456	\\
39.7163	&	0.271	&	39.7167	&	-0.145	&	39.7172	&	-0.419	&	39.7176	&	-0.665	&	39.8434	&	0.258	&	39.8438	&	-0.164	&	39.8443	&	-0.450	\\
39.7180	&	0.263	&	39.7184	&	-0.142	&	39.7188	&	-0.438	&	39.7192	&	-0.663	&	39.8452	&	0.281	&	39.8457	&	-0.166	&	39.8461	&	-0.430	\\
39.7197	&	0.254	&	39.7202	&	-0.167	&	39.7205	&	-0.448	&	39.7210	&	-0.656	&	39.8470	&	0.306	&	39.8474	&	-0.154	&	39.8480	&	-0.440	\\
39.7215	&	0.256	&	39.7220	&	-0.172	&	39.7226	&	-0.425	&	39.7233	&	-0.686	&	39.8489	&	0.274	&	39.8493	&	-0.145	&	39.8497	&	-0.399	\\
39.7241	&	0.250	&	39.7247	&	-0.189	&	39.7251	&	-0.444	&	39.7255	&	-0.681	&	39.8506	&	0.294	&	39.8511	&	-0.126	&	39.8516	&	-0.393	\\
39.7260	&	0.256	&	39.7264	&	-0.167	&	39.7269	&	-0.455	&	39.7274	&	-0.676	&	39.8525	&	0.305	&	39.8529	&	-0.114	&	39.8534	&	-0.435	\\
39.7279	&	0.254	&	39.7284	&	-0.181	&	39.7289	&	-0.439	&	39.7293	&	-0.698	&	39.8544	&	0.307	&	39.8548	&	-0.118	&	40.7574	&	-0.179	\\
39.7298	&	0.236	&	39.7302	&	-0.173	&	39.7308	&	-0.452	&	39.7312	&	-0.705	&	40.7582	&	0.542	&	40.7587	&	0.101	&	40.7590	&	-0.167	\\
39.7318	&	0.243	&	39.7322	&	-0.167	&	39.7326	&	-0.445	&	39.7331	&	-0.695	&	40.7599	&	0.561	&	40.7603	&	0.108	&	40.7607	&	-0.178	\\
39.7336	&	0.241	&	39.7343	&	-0.194	&	39.7348	&	-0.461	&	39.7352	&	-0.687	&	40.7615	&	0.573	&	40.7619	&	0.128	&	40.7623	&	-0.130	\\
39.7356	&	0.242	&	39.7361	&	-0.189	&	39.7364	&	-0.464	&	39.7368	&	-0.698	&	40.7632	&	0.575	&	40.7636	&	0.129	&	40.7640	&	-0.141	\\
39.7373	&	0.235	&	39.7377	&	-0.189	&	39.7381	&	-0.444	&	39.7385	&	-0.687	&	40.7649	&	0.586	&	40.7653	&	0.133	&	40.7656	&	-0.132	\\
39.7390	&	0.226	&	39.7394	&	-0.190	&	39.7398	&	-0.452	&	39.7402	&	-0.686	&	40.7665	&	0.587	&	40.7669	&	0.126	&	40.7674	&	-0.140	\\
39.7407	&	0.228	&	39.7411	&	-0.197	&	39.7415	&	-0.453	&	39.7419	&	-0.709	&	40.7683	&	0.587	&	40.7687	&	0.130	&	40.7690	&	-0.108	\\
39.7423	&	0.235	&	39.7427	&	-0.186	&	39.7431	&	-0.474	&	39.7436	&	-0.706	&	40.7699	&	0.599	&	40.7703	&	0.148	&	40.7707	&	-0.120	\\
39.7440	&	0.229	&	39.7443	&	-0.189	&	39.7447	&	-0.471	&	39.7451	&	-0.692	&	40.7716	&	0.591	&	40.7720	&	0.129	&	40.7724	&	-0.111	\\
39.7456	&	0.231	&	39.7460	&	-0.209	&	39.7464	&	-0.474	&	39.7468	&	-0.703	&	40.7733	&	0.603	&	40.7737	&	0.154	&	40.7740	&	-0.129	\\
39.7472	&	0.225	&	39.7476	&	-0.202	&	39.7480	&	-0.481	&	39.7484	&	-0.688	&	40.7749	&	0.601	&	40.7753	&	0.154	&	40.7757	&	-0.102	\\
39.7489	&	0.212	&	39.7493	&	-0.205	&	39.7497	&	-0.477	&	39.7501	&	-0.717	&	40.7766	&	0.603	&	40.7770	&	0.153	&	40.7774	&	-0.108	\\
39.7505	&	0.224	&	39.7509	&	-0.204	&	39.7513	&	-0.467	&	39.7517	&	-0.721	&	40.7783	&	0.611	&	40.7787	&	0.150	&	40.7791	&	-0.103	\\
39.7522	&	0.215	&	39.7526	&	-0.197	&	39.7530	&	-0.474	&	39.7535	&	-0.699	&	40.7800	&	0.618	&	40.7804	&	0.168	&	40.7808	&	-0.116	\\
\hline
\end{tabular}
\end{tiny}
\end{center}
\end{table}

The coordinates of the comparison and check stars are listed in Table 1. The corresponding light curves are displayed in Fig. 1 where the magnitude differences between the comparison star and the check star are shown in the figure. The second set of light curve were obtained from March to April 2017 and plotted in Fig. 2. To obtain more times of light minimum, TY Pup was also monitored by using the 60 cm and 1.0-m telescopes of Yunnan Observatories (YNOs) in February 2015 and January 2018, respectively. These telescopes were equipped with a Cassegrain-focus multicolor CCD photometer where an Andor DW436\,2K CCD camera. Standard Johnson-Cousin-Bessel $BV(RI)_C$ filters were used. The eclipse profiles obtained from Yunnan Observatories are shown in Fig. 3. The two sets of photometric data for TY Pup in magnitude differences between the variable star and the comparison star with heliocentric Julian dates are listed in the online Table 2 and 3 for 2015 and 2017, respectively.

\begin{table}
\caption{CCD observations in BVRI bands for TY Pup observed in 2017}
\begin{center}
\begin{tiny}
\begin{tabular}{cccccccccccccc}\hline\hline
HJD&$\Delta B$ &HJD &$\Delta V$ &HJD &$\Delta R$ &HJD&$\Delta I$&HJD&$\Delta B$ &HJD &$\Delta V$ &HJD &$\Delta R$ \\
-2457800&mag&&&&&&&&&&&&\\
\hline
33.6422	&	0.203	&	33.6425	&	-0.238	&	33.6427	&	-0.506	&	33.6430	&	-0.715	&	35.5474	&	0.466	&	35.5527	&	-0.009	&	35.5530	&	-0.276	\\
33.6439	&	0.219	&	33.6441	&	-0.249	&	33.6444	&	-0.502	&	33.6448	&	-0.710	&	35.5489	&	0.481	&	35.5542	&	-0.014	&	35.5545	&	-0.275	\\
33.6512	&	0.210	&	33.6515	&	-0.256	&	33.6517	&	-0.464	&	33.6520	&	-0.720	&	35.5525	&	0.453	&	35.5557	&	-0.069	&	35.5560	&	-0.301	\\
33.6530	&	0.221	&	33.6532	&	-0.222	&	33.6552	&	-0.488	&	33.6537	&	-0.697	&	35.5539	&	0.441	&	35.5571	&	-0.052	&	35.5574	&	-0.299	\\
33.6546	&	0.224	&	33.6549	&	-0.230	&	33.6571	&	-0.478	&	33.6555	&	-0.721	&	35.5553	&	0.432	&	35.5589	&	-0.029	&	35.5592	&	-0.304	\\
33.6565	&	0.203	&	33.6569	&	-0.235	&	33.6588	&	-0.463	&	33.6574	&	-0.722	&	35.5568	&	0.441	&	35.5604	&	-0.056	&	35.5607	&	-0.292	\\
33.6584	&	0.229	&	33.6586	&	-0.229	&	33.6618	&	-0.491	&	33.6591	&	-0.691	&	35.5586	&	0.436	&	35.5622	&	-0.020	&	35.5625	&	-0.304	\\
33.6611	&	0.232	&	33.6614	&	-0.214	&	33.6638	&	-0.471	&	33.6621	&	-0.692	&	35.5600	&	0.412	&	35.5636	&	-0.068	&	35.5639	&	-0.284	\\
33.6631	&	0.241	&	33.6634	&	-0.240	&	33.6670	&	-0.460	&	33.6641	&	-0.697	&	35.5618	&	0.418	&	35.5650	&	-0.069	&	35.5653	&	-0.317	\\
33.6664	&	0.233	&	33.6667	&	-0.222	&	33.6688	&	-0.464	&	33.6673	&	-0.701	&	35.5632	&	0.414	&	35.5664	&	-0.068	&	35.5666	&	-0.328	\\
33.6683	&	0.237	&	33.6686	&	-0.240	&	33.6707	&	-0.473	&	33.6691	&	-0.716	&	35.5646	&	0.419	&	35.5679	&	-0.075	&	35.5682	&	-0.323	\\
33.6703	&	0.245	&	33.6705	&	-0.197	&	33.6960	&	-0.439	&	33.6710	&	-0.694	&	35.5661	&	0.402	&	35.5697	&	-0.075	&	35.5700	&	-0.333	\\
33.6954	&	0.290	&	33.6957	&	-0.180	&	33.6978	&	-0.398	&	33.6962	&	-0.653	&	35.5675	&	0.405	&	35.5781	&	-0.087	&	35.5785	&	-0.359	\\
33.6972	&	0.295	&	33.6976	&	-0.176	&	33.6997	&	-0.384	&	33.6980	&	-0.670	&	35.5693	&	0.404	&	35.5797	&	-0.098	&	35.5800	&	-0.359	\\
33.6990	&	0.293	&	33.6994	&	-0.167	&	34.5324	&	-0.390	&	33.6999	&	-0.652	&	35.5729	&	0.383	&	35.5815	&	-0.088	&	35.5818	&	-0.336	\\
33.7009	&	0.285	&	33.7011	&	-0.125	&	34.5338	&	-0.374	&	34.5326	&	-0.625	&	35.5794	&	0.374	&	35.5829	&	-0.108	&	35.5831	&	-0.369	\\
34.5331	&	0.317	&	34.5334	&	-0.127	&	34.5352	&	-0.384	&	34.5340	&	-0.612	&	35.5811	&	0.367	&	35.5842	&	-0.105	&	35.5846	&	-0.381	\\
34.5345	&	0.313	&	34.5349	&	-0.118	&	34.5367	&	-0.399	&	34.5356	&	-0.619	&	35.5826	&	0.357	&	35.5858	&	-0.128	&	35.5861	&	-0.383	\\
34.5360	&	0.323	&	34.5363	&	-0.153	&	34.5380	&	-0.377	&	34.5369	&	-0.604	&	35.5840	&	0.372	&	35.5872	&	-0.111	&	35.5874	&	-0.372	\\
34.5374	&	0.327	&	34.5377	&	-0.108	&	34.5394	&	-0.362	&	34.5383	&	-0.592	&	35.5854	&	0.347	&	35.5886	&	-0.121	&	35.5890	&	-0.381	\\
34.5389	&	0.332	&	34.5391	&	-0.108	&	34.5409	&	-0.376	&	34.5398	&	-0.591	&	35.5869	&	0.363	&	35.5900	&	-0.131	&	35.5903	&	-0.378	\\
34.5403	&	0.345	&	34.5405	&	-0.128	&	34.5424	&	-0.353	&	34.5411	&	-0.598	&	35.5883	&	0.373	&	35.5935	&	-0.117	&	35.5939	&	-0.396	\\
34.5417	&	0.338	&	34.5421	&	-0.113	&	34.5436	&	-0.375	&	34.5428	&	-0.602	&	35.5897	&	0.353	&	35.5951	&	-0.127	&	35.5954	&	-0.388	\\
34.5432	&	0.348	&	34.5434	&	-0.107	&	34.5465	&	-0.381	&	34.5440	&	-0.596	&	35.5912	&	0.337	&	35.5965	&	-0.123	&	35.5969	&	-0.393	\\
34.5459	&	0.359	&	34.5462	&	-0.109	&	34.5492	&	-0.359	&	34.5495	&	-0.591	&	35.5947	&	0.328	&	35.5980	&	-0.132	&	35.5983	&	-0.406	\\
34.5486	&	0.354	&	34.5490	&	-0.091	&	34.5540	&	-0.349	&	34.5543	&	-0.569	&	35.5962	&	0.326	&	35.5998	&	-0.133	&	35.6001	&	-0.394	\\
34.5514	&	0.373	&	34.5536	&	-0.092	&	34.5569	&	-0.317	&	34.5572	&	-0.564	&	35.5976	&	0.325	&	35.6012	&	-0.150	&	35.6015	&	-0.399	\\
34.5562	&	0.381	&	34.5566	&	-0.068	&	34.5597	&	-0.309	&	34.5599	&	-0.539	&	35.5994	&	0.326	&	35.6026	&	-0.136	&	35.6029	&	-0.415	\\
34.5591	&	0.388	&	34.5593	&	-0.080	&	34.5623	&	-0.315	&	34.5627	&	-0.546	&	35.6008	&	0.316	&	35.6041	&	-0.152	&	35.6043	&	-0.406	\\
34.5618	&	0.395	&	34.5621	&	-0.042	&	34.5650	&	-0.326	&	34.5653	&	-0.557	&	35.6023	&	0.317	&	35.6054	&	-0.160	&	35.6057	&	-0.421	\\
34.5645	&	0.406	&	34.5648	&	-0.068	&	34.5678	&	-0.309	&	34.5681	&	-0.521	&	35.6037	&	0.321	&	35.6069	&	-0.153	&	35.6071	&	-0.423	\\
34.5672	&	0.414	&	34.5703	&	-0.020	&	34.5707	&	-0.304	&	34.5709	&	-0.523	&	35.6051	&	0.330	&	35.6084	&	-0.149	&	35.6087	&	-0.409	\\
34.5700	&	0.426	&	34.5750	&	-0.033	&	34.5752	&	-0.280	&	34.5755	&	-0.486	&	35.6066	&	0.305	&	35.6098	&	-0.156	&	35.6101	&	-0.409	\\
34.5727	&	0.440	&	34.5791	&	-0.001	&	34.5793	&	-0.260	&	34.5797	&	-0.488	&	35.6080	&	0.303	&	35.6116	&	-0.172	&	35.6119	&	-0.418	\\
34.5788	&	0.442	&	34.5819	&	0.035	&	34.5821	&	-0.254	&	34.5823	&	-0.496	&	35.6094	&	0.300	&	35.6149	&	-0.169	&	35.6151	&	-0.433	\\
34.5815	&	0.461	&	34.5916	&	0.059	&	34.5918	&	-0.208	&	34.5920	&	-0.465	&	35.6112	&	0.292	&	35.6162	&	-0.167	&	35.6166	&	-0.436	\\
34.5912	&	0.510	&	34.6098	&	0.129	&	34.6101	&	-0.125	&	34.6105	&	-0.369	&	35.6127	&	0.299	&	35.6180	&	-0.174	&	35.6183	&	-0.443	\\
34.5939	&	0.520	&	34.6153	&	0.150	&	34.6129	&	-0.122	&	34.6131	&	-0.342	&	35.6159	&	0.291	&	35.6194	&	-0.181	&	35.6198	&	-0.430	\\
34.6123	&	0.594	&	34.6200	&	0.160	&	34.6174	&	-0.113	&	34.6177	&	-0.352	&	35.6177	&	0.282	&	35.6209	&	-0.165	&	35.6212	&	-0.447	\\
34.6149	&	0.600	&	34.6364	&	0.160	&	34.6202	&	-0.108	&	34.6204	&	-0.333	&	35.6191	&	0.279	&	35.6223	&	-0.174	&	35.6227	&	-0.449	\\
34.6175	&	0.607	&	34.6410	&	0.147	&	34.6385	&	-0.095	&	34.6444	&	-0.311	&	35.6205	&	0.276	&	35.6241	&	-0.183	&	35.6243	&	-0.452	\\
34.6361	&	0.610	&	34.6437	&	0.132	&	34.6413	&	-0.109	&	34.6470	&	-0.328	&	35.6220	&	0.272	&	35.6256	&	-0.206	&	35.6259	&	-0.444	\\
34.6406	&	0.615	&	34.6465	&	0.165	&	34.6441	&	-0.125	&	34.6514	&	-0.324	&	35.6238	&	0.265	&	35.6273	&	-0.201	&	35.6277	&	-0.448	\\
34.6434	&	0.608	&	34.6508	&	0.158	&	34.6468	&	-0.095	&	34.6541	&	-0.323	&	35.6252	&	0.274	&	35.6288	&	-0.196	&	35.6291	&	-0.462	\\
34.6462	&	0.607	&	34.6536	&	0.171	&	34.6510	&	-0.093	&	35.5322	&	-0.406	&	35.6270	&	0.272	&	35.6302	&	-0.183	&	35.6305	&	-0.446	\\
34.6505	&	0.612	&	34.6563	&	0.155	&	34.6538	&	-0.087	&	35.5335	&	-0.420	&	35.6284	&	0.263	&	35.6316	&	-0.202	&	35.6320	&	-0.463	\\
34.6532	&	0.608	&	35.5301	&	0.076	&	35.5303	&	-0.183	&	35.5365	&	-0.454	&	35.6299	&	0.264	&	35.6330	&	-0.210	&	35.6333	&	-0.460	\\
34.6560	&	0.623	&	35.5316	&	0.087	&	35.5319	&	-0.168	&	35.5380	&	-0.443	&	35.6313	&	0.256	&	35.6364	&	-0.195	&	35.6367	&	-0.470	\\
35.5298	&	0.536	&	35.5330	&	0.075	&	35.5333	&	-0.198	&	35.5393	&	-0.460	&	35.6327	&	0.266	&	35.6381	&	-0.216	&	35.6384	&	-0.483	\\
35.5313	&	0.539	&	35.5345	&	0.073	&	35.5349	&	-0.212	&	35.5409	&	-0.440	&	35.6360	&	0.251	&	35.6399	&	-0.215	&	35.6402	&	-0.485	\\
35.5327	&	0.535	&	35.5358	&	0.060	&	35.5362	&	-0.221	&	35.5422	&	-0.454	&	35.6378	&	0.251	&	35.6423	&	-0.198	&	35.6426	&	-0.453	\\
35.5342	&	0.524	&	35.5374	&	0.034	&	35.5377	&	-0.204	&	35.5437	&	-0.472	&	35.6396	&	0.240	&	35.6445	&	-0.234	&	35.6449	&	-0.461	\\
35.5356	&	0.526	&	35.5387	&	0.023	&	35.5391	&	-0.208	&	35.5450	&	-0.472	&	35.6419	&	0.249	&	35.6460	&	-0.229	&	35.6462	&	-0.476	\\
35.5371	&	0.504	&	35.5403	&	0.030	&	35.5406	&	-0.223	&	35.5467	&	-0.476	&	35.6443	&	0.231	&	35.6476	&	-0.226	&	35.6479	&	-0.483	\\
35.5385	&	0.502	&	35.5417	&	0.035	&	35.5419	&	-0.217	&	35.5484	&	-0.494	&	35.6456	&	0.229	&	35.6494	&	-0.236	&	35.6498	&	-0.474	\\
35.5399	&	0.510	&	35.5430	&	0.028	&	35.5434	&	-0.240	&	35.5518	&	-0.475	&	35.6474	&	0.233	&	35.6513	&	-0.224	&	35.6516	&	-0.499	\\
35.5414	&	0.486	&	35.5446	&	0.018	&	35.5448	&	-0.235	&	35.5534	&	-0.489	&	35.6491	&	0.219	&	35.6531	&	-0.225	&	35.6535	&	-0.475	\\
35.5428	&	0.487	&	35.5460	&	0.040	&	35.5464	&	-0.244	&	35.5548	&	-0.508	&	35.6509	&	0.225	&	35.6550	&	-0.218	&	35.6553	&	-0.473	\\
35.5442	&	0.488	&	35.5478	&	0.017	&	35.5481	&	-0.246	&	35.5563	&	-0.500	&	35.6528	&	0.230	&	35.6567	&	-0.252	&	35.6571	&	-0.506	\\
35.5457	&	0.455	&	35.5511	&	-0.001	&	35.5514	&	-0.277	&	35.5578	&	-0.522	&	35.6546	&	0.223	&	35.6585	&	-0.227	&	35.6588	&	-0.497	\\
\hline
\end{tabular}
\end{tiny}
\end{center}
\end{table}

\begin{figure}
\begin{center}
\includegraphics[angle=0,scale=0.4]{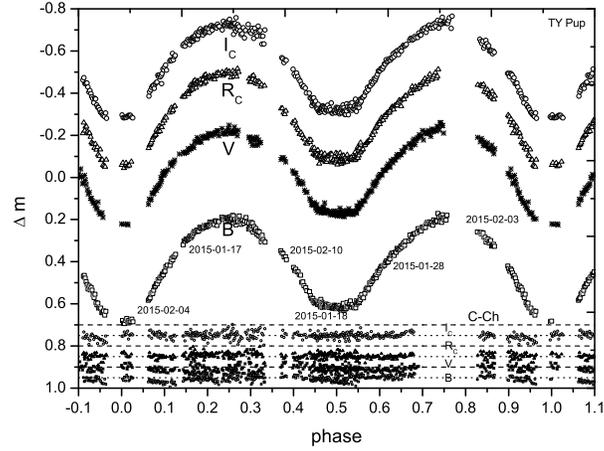}
\caption{Multi-color CCD light curves in $B$, $V$, $(RI)_c$ bands
obtained with the 0.6-m telescope at CTIO in January and February 2015.
The differential magnitudes between the comparison and the check stars are also presented.}
\end{center}
\end{figure}

\begin{figure}
\begin{center}
\includegraphics[angle=0,scale=0.4]{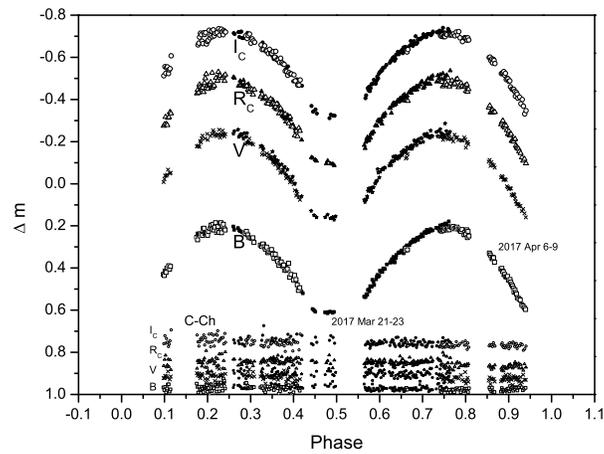}
\caption{CCD photometric observations in $B$, $V$, $(RI)_c$ bands were
obtained with the 0.6-m telescope at CTIO in March and April 2017. As those shown in Fig. 1, the magnitude differences between the comparison and the check stars are shown.}
\end{center}
\end{figure}

\begin{figure}
\begin{center}
\includegraphics[angle=0,scale=0.4]{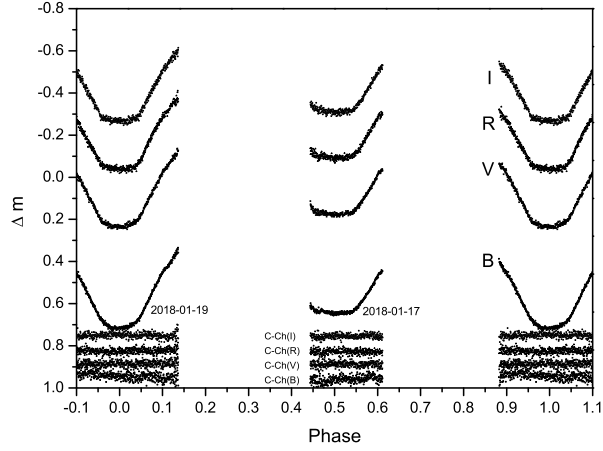}
\caption{Eclipse profiles in $B$, $V$, $R$ and $I$ bands were
obtained with the 1.0-m telescope at YNOs in January 2018.}
\end{center}
\end{figure}

\section{Variations in the orbital period}

Earlier epochs and $O-C$ analyses of TY Pup were published by several investigators (e.g. Gu et al. 1993 and Berdnikov \& Turner 1995). The authors derived a linear ephemeris for the binary. Later, Qian (2001a) obtained a quadratic ephemeris and pointed out that the period of TY Pup was secular increasing with rate of $\mathrm{d}P /\mathrm{d}t=1.66\times 10^{-7}$ d $\textrm{yr}^{-1}$ and $\dot{P}/P$ = 2.03$\times 10^{-7}$ $\textrm{yr}^{-1}$.

\begin{figure}
\begin{center}
\includegraphics[angle=0,scale=0.4]{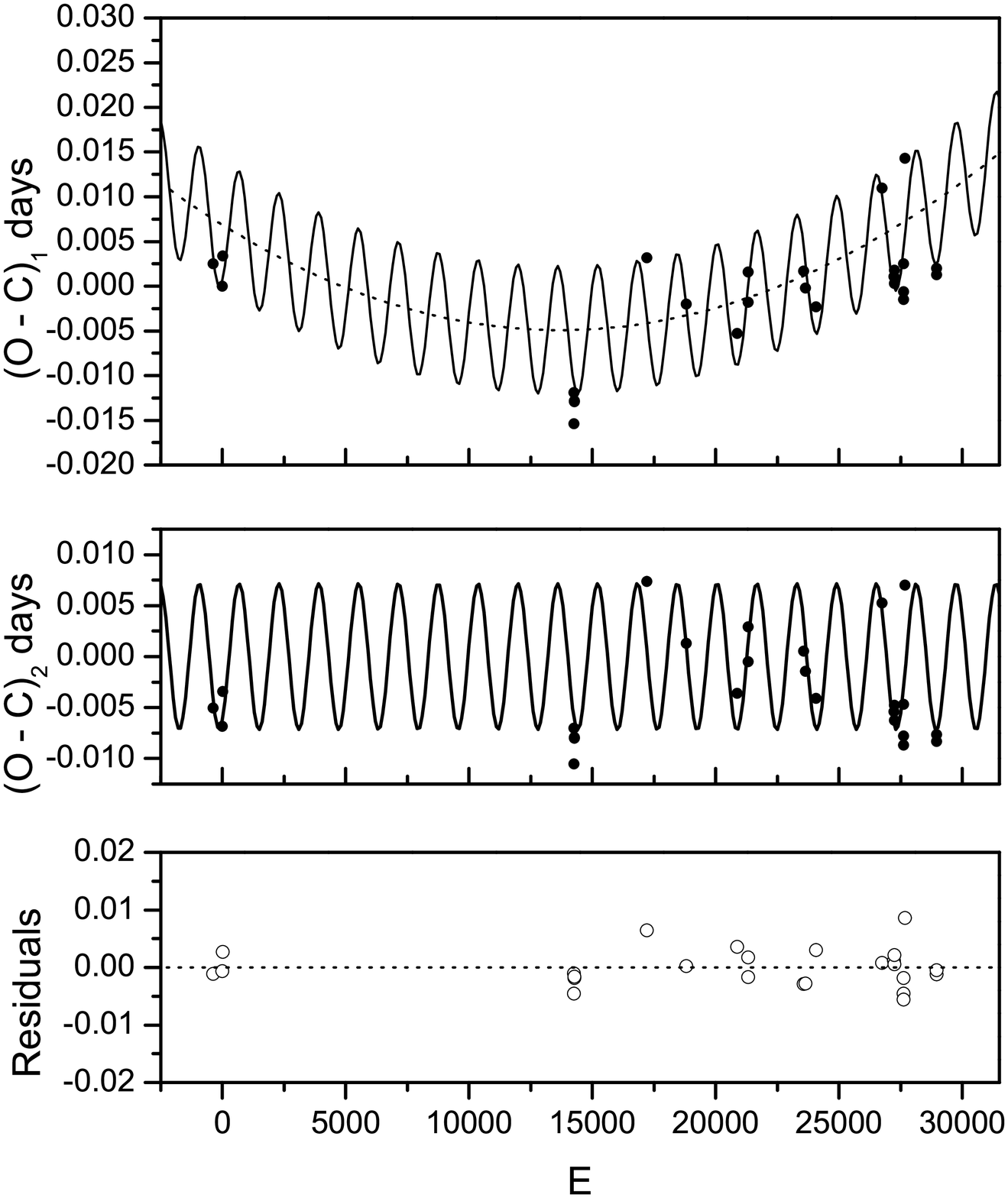}
\caption{The $(O-C)_1$ diagram was constructed by using the linear
ephemeris in Eq. (1). The solid line in the upper panel refers to a combination of a long-term period increase and a semi-amplitude cyclic oscillation, while the dashed line refers to the long-term increase.}
\end{center}
\end{figure}

Based on our photometric observations, four times of light minimum were determined. All times of minimum light are listed in Table 4. The variations of the orbital period were analysed by using $O-C$ (observed minus calculated) method. In order to investigate the orbital period change of TY Pup, the $(O-C)_1$ values of all available times of light minimum were computed with the linear ephemeris given by Kreiner (2004):
\begin{equation}
Min.I (HJD) = 2434412.106+0^{d}.8192423\times{E}.
\end{equation}
The corresponding O-C diagram is shown in the upper panel of Fig. 4. As shown in the panel, the changes in the orital period of TY Pup are complex due to a small-amplitude cyclic variation and an upward parabolic variation cannot fit the $(O-C)_1$ curve very well. To get a better fit for the trend of $(O-C)_1$ curve, we have to combine a new quadratic ephemeris with an additional sinusoidal term. By using a least-squares method, the new ephemeris was determined:
\begin{equation}
\begin{array}{lll}
Min.I(HJD) =  2434412.1128(\pm 0.0002) +0.8192406(\pm 0.0000003)E
\\ +[62.5(\pm 0.9)\times 10^{-12}]E^2
\\ +0.0072(\pm 0.0001) \times \sin[0.^{\circ}2229E+60.^{\circ}0(\pm 0.^{\circ}9)].
\end{array}
\end{equation}
According to Eq. (2), the semi-amplitude of cyclic oscillation is 0.0072 days and the sinusoidal term suggests an oscillation period of 3.62 years. The quadratic term in Eq. (2) also reveals a continuous period increase at a rate of $\mathrm{d}P/\mathrm{d}t = 5.57(\pm 0.08)\times10^{-8}$ d $\textrm{yr}^{-1}$. This kind of period variation is usually encountered for W UMa-type binary stars. Some other examples are AB And and TY UMa (e.g., Li et al. 2014, 2015). After the long-term period change is subtracted from the O-C diagram, the cyclic oscillation is shown in the middle panel of Fig. 4. The residuals of Eq. (2) are plotted in the lowest panel. However, there are no times of light minimum recorded between E=0 and E=14250, thus new eclipse times are required in the future to confirm the variations presented here.

\begin{table}
\scriptsize
\caption{Times of minimum light for TY Pup.}
\begin{center}
\begin{tabular}{lcrrccc}\hline\hline
HJD(2400000+) & Error(days) & E & $(O-C)_1$ & Method & Min & Ref. \\
\hline
34092.6040 & 0.0009 & -390.0  & 0.0025  & pe &  I & (1)\\
34412.1060 &        & 0.0     & 0.0000  & pe &  I & (4)\\
34416.2056 & 0.0008 & 5.0     & 0.0034  & pe &  I & (1)\\
46086.2934 &        & 14250.0 & -0.0154 & pe &  I & (2)\\
46087.1161 &        & 14251.0 & -0.0119 & pe &  I & (2)\\
46100.2230 & 0.0002 & 14267.0 & -0.0129 & pe &  I & (2)\\
46107.1867 &        & 14275.5 & -0.0128 & pe & II & (2)\\
48500.6190 &        & 17197.0 & 0.0032  & ccd & I & (5)\\
49817.1362 & 0.0013 & 18804.0 & -0.0020 & pe &  I & (3)\\
51508.0490 &        & 20868.0 & -0.0053 & ccd & I & (5)\\
51868.9321 & 0.0013 & 21308.5 & 0.0016  & ccd & II& (6)\\
51869.3383 & 0.0013 & 21309.0 & -0.0018 & ccd & I & (6)\\
53714.2755 &        & 23561.0 & 0.0017  & ccd & I & (5)\\
53778.9938 &        & 23640.0 & -0.0002 & ccd & I & (5)\\
54120.6157 &        & 24057.0 & -0.0023 & ccd & I & (5)\\
56323.9812 &        & 26746.5 & 0.0110  & ccd & II& (5)\\
56714.3402 & 0.0003 & 27223.0 & 0.0011  & ccd & I & (7)\\
56730.3147 & 0.0005 & 27242.5 & 0.0003  & ccd & II& (7)\\
56737.2797 & 0.0003 & 27251.0 & 0.0018  & ccd & I & (7)\\
57028.5180 &        & 27606.5 & -0.0006 & ccd & II& (5)\\
57033.0229 &        & 27612.0 & -0.0015 & ccd & I & (5)\\
57040.8097 & 0.0013 & 27621.5 & 0.0025	& ccd & II& PROMPT-8\\
57080.1451 & 0.0004 & 27669.5 & 0.0143  & ccd & II& YNOs 60-cm\\
58136.1354 & 0.0006 & 28958.5 & 0.0013  & ccd & II& YNOs 1.0-m\\
58138.1842 & 0.0002 & 28961.0 & 0.0020  & ccd & I & YNOs 1.0-m\\
\hline
\end{tabular}
\end{center}
{\footnotesize References.} \footnotesize (1) Huruhata et al. 1957; (2) Gu et al. 1993; (3) Berdnikov \& Turner 1995; (4) Van Houten 1971; (5) http://var.astro.cz/ocgate; (6) Pojmanski 1997 \& 2002; (7) Karampotsiou et al. 2016.
\end{table}

\section{Photometric solutions with W-D method}

The light curve of W UMa-type binary stars are usually varying with time.
Some examples with variable light curves are FG Hya (Qian \& Yang 2005), AD
Cnc (Qian et al. 2007), BX Peg (Lee et al. 2004), CU Tau (Qian et al. 2005), CE Leo (Kang et al. 2004), EQ Tau (Yuan \& Qian, 2007), and QX And (Qian et al. 2007). To check whether the light curve of TY Pup is variable or not, we compare our light curves obtained in 2015 and 2017 as shown in Fig. 5, the light curves generally overlap within the error and all of them are clearly symmetric indicating that the light curve may not be variable.

\begin{figure}
\begin{center}
\includegraphics[angle=0,scale=0.5]{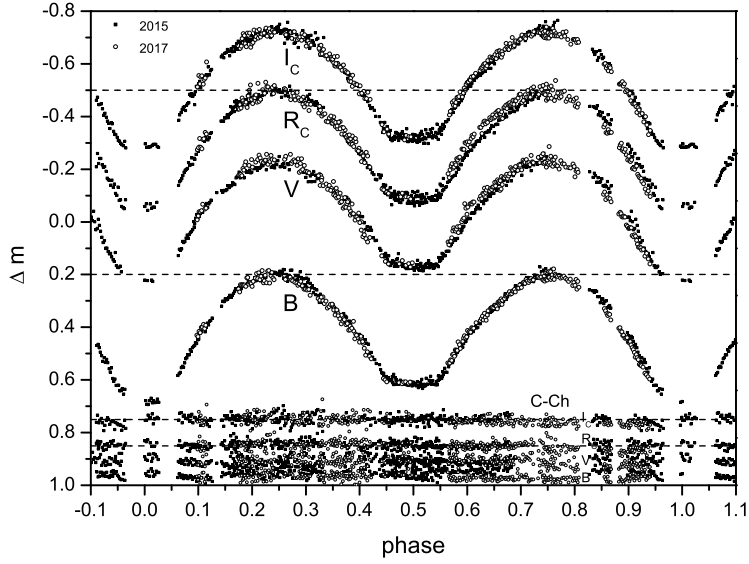}
\caption{The comparison between our light curves obtained during 2015 and 2017 in $B$, $V$, $R_C$ and $I_C$ bands.}
\end{center}
\end{figure}

{\bf
\begin{figure}
\begin{center}
\includegraphics[angle=0,scale=1]{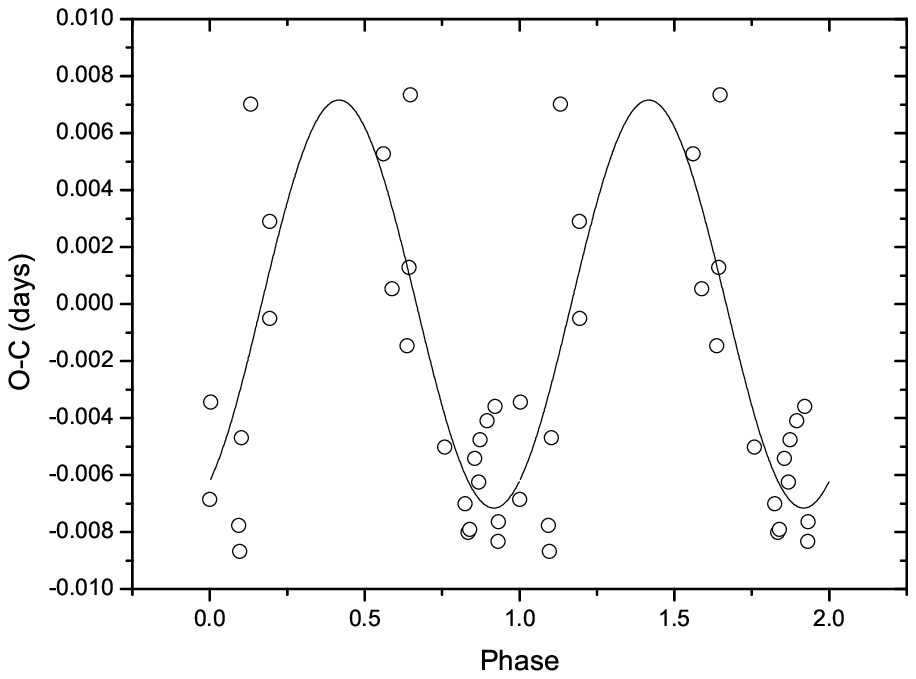}
\caption{The variation in orbital period from middle panel of Fig. 4 as phase scale. The computed curve almost covers to all $O-C$ data, the maxima and the minima are quite constrained to the curve, indicating that the period change is reliable.}
\end{center}
\end{figure}
}

For the photometric solution, we use the spectral type of F3V determined by Duerbeck \& Rucinski (2007). Our photometric data in four-color $BV(RI)_C$ light curves observed in 2015 are analysed by using the Wilson \& Devinney (W-D) code (Wilson \& Devinney 1971; Wilson 1990, 1994, 2012; van Hamme \& Wilson 2007) to determine their photometric elements. The color index $B-V$=0.40 given by Morton \& Adams (1968) corresponds to $T_{eff}$=7000\,K, while the value $B-V$=0.39 where $T_{eff}$=6900\,K (Flower 1996). During the solutions, the effective temperature of the primary star ($T_1$) was fixed as 6900 K corresponding to its spectral type (Cox 2000). We assume that the convective envelope is already developed for both components. Therefore, the bolometric albedos for star 1 and 2 were taken as $A_{1} = A_{2}$ = 0.5 (Rucinski 1969) and the values of the gravity-darkening coefficients $g_{1} = g_{2}$ = 0.32 (Lucy 1967) were used. The monochromatic and bolometric limb-darkening coefficients were logarithmically interpolated from van Hamme's table (van Hamme 1993).

\begin{figure}
\begin{center}
\includegraphics[angle=0,scale=0.4]{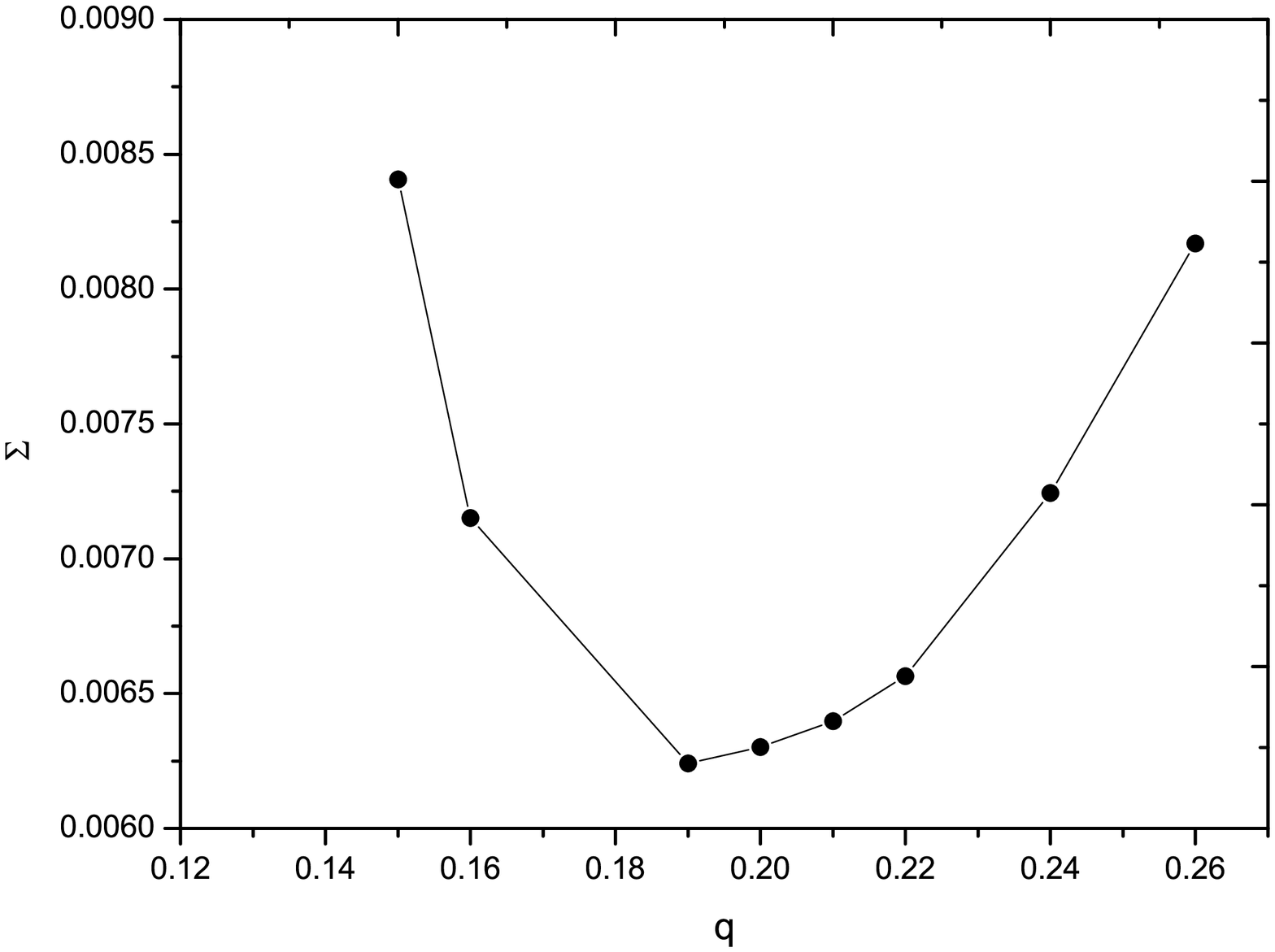}
\caption{Relation between the mass ratio $q$ and the sum of weighted square deviations ($\Sigma$). The graph indicates that the optimal mass ratio $q$ is in the range between 0.18 and 0.22.}
\end{center}
\end{figure}

\begin{figure}
\begin{center}
\includegraphics[angle=0,scale=0.4]{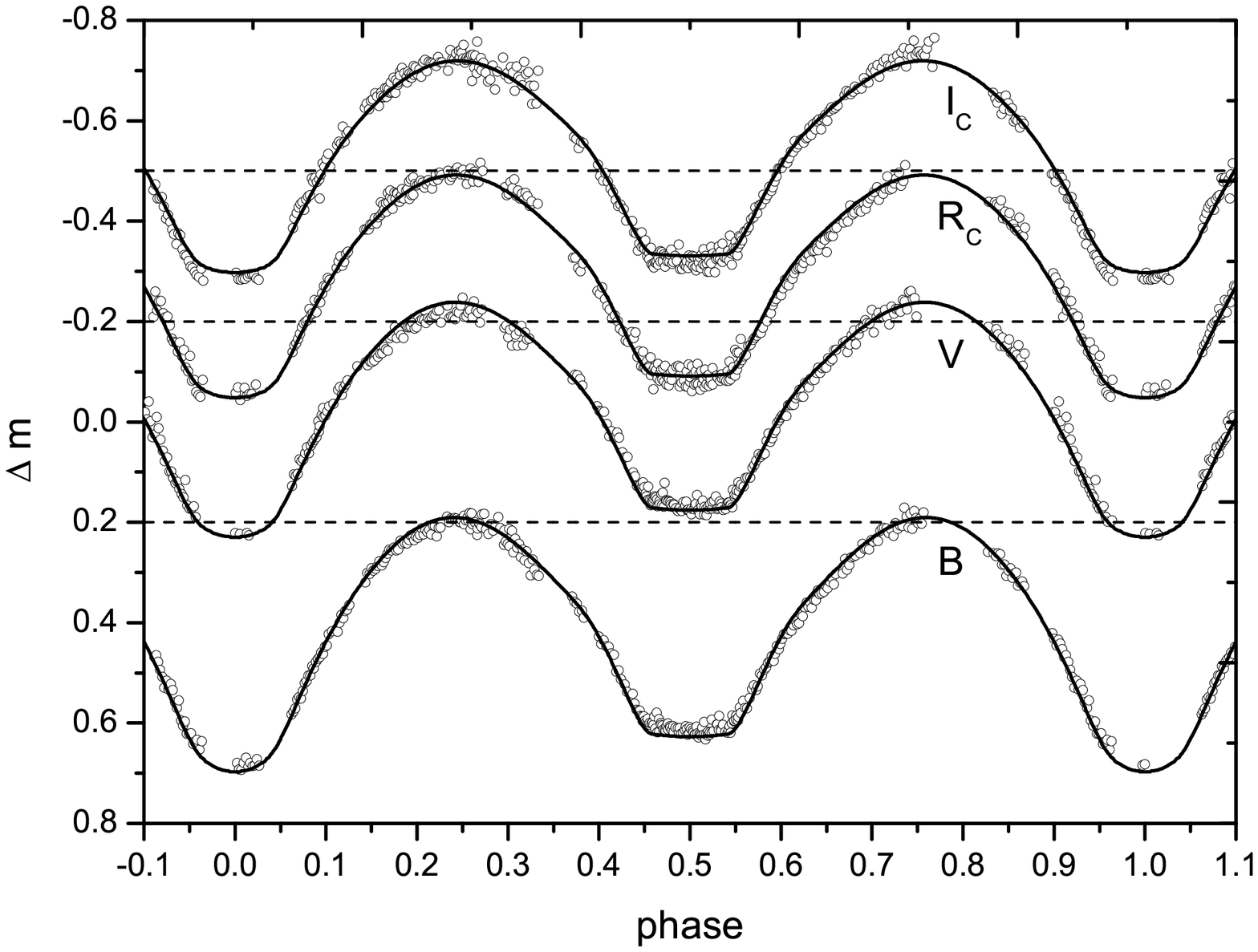}
\caption{Theoretical light curves (solid lines) calculated using the W-D method compare to the normal points of the observed light curves from photometric data in January-February 2015. }
\end{center}
\end{figure}

Ideally, for reliable masses the mass ratio should be obtained from precise spectroscopic radial velocity measurements (Deb \& Singh 2011). But for our photometric data, we found that the synthetic light curves could not fit well when we used the spectroscopic mass ratio $q_{sp}$ of 0.25$\pm 0.03$ from Duerbeck \& Rucinski (2007). Therefore, We used a $q$-search method to determine its photometric mass ratio $q_{ph}$ and then set the mass ratio as an adjustable parameter to get a better fit. The $q$-search result suggests that the range of mass ratio is between 0.18 to 0.22 as displayed in Fig. 7. At the end of modeling process, we obtained the photometric mass ratio of 0.1839($\pm0.0016$) at the lowest sum of the weighted square deviations $\Sigma (\omega(O-C))^2$ or hereafter $\Sigma$.

The adjustable parameters are the inclination ($i$), the mass ratio ($q$), the temperature of Star 2 ($T_{2}$), the monochromatic luminosity of Star 1 ($L_{1B}$, $L_{1V}$, $L_{1R}$ and $L_{1I}$), the dimensionless potential of stars 1 ($\Omega_{1}=\Omega_{2}$) in mode 3 (Leung \& Willson 1977) for contact configuration, respectively. As shown in Fig. 5, the light curves in $BV(RI)_C$ bands seem to be symmetric, so no spot model was considered. In addition, the $(O-C)$ diagram shows a cyclic variation that may be caused by light-travel time effect via the presence of a third companion. Thus, we added the third light ($l_{3}$) as an adjustable parameter in the modeling process to get a better fit. As the result, the third light could not be detected during the process. It seems to be very small contribution when compared to the total light from the system. The solutions are listed in Table 5 and theoretical light curves (solid lines) are plotted in Fig. 8, compared to the normal points from photometric observations.

\begin{table}
\scriptsize
\caption{Photometric solutions with formal errors.}
\begin{center}
\begin{tabular}{lcccc}\hline\hline
Parameters                   & Data (2015)\\
\hline
$T_1(K)$            & 6900 (fixed)\\
$g_1=g_2$           & 0.32 (fixed)\\
$A_1=A_2$           & 0.50 (fixed)\\
$q$                 &0.1839($\pm0.0016$) \\
$T_2(K)$            &6915($\pm10$) \\
$i(^o)$             &83.638($\pm0.189$) \\
$\Omega_{in}$       &2.1923 \\
$\Omega_{out}$      &2.0745 \\
$\Omega_1=\Omega_2$ &2.0930($\pm0.0049$)\\
$L_1/(L_1+L_2)$(B)  &0.7982($\pm0.0015$)\\
$L_1/(L_1+L_2)$(V)  &0.7996($\pm0.0011$)\\
$L_1/(L_1+L_2)$(R)  &0.8004($\pm0.0009$)\\
$L_1/(L_1+L_2)$(I)  &0.8010($\pm0.0009$)\\
$r_1(pole)$         &0.5181($\pm0.0009$)\\
$r_1(side)$         &0.5753($\pm0.0014$)\\
$r_1(back)$         &0.6060($\pm0.0014$)\\
$r_2(pole)$         &0.2568($\pm0.0044$)\\
$r_2(side)$         &0.2720($\pm0.0056$)\\
$r_2(back)$         &0.3564($\pm0.0260$)\\
$f$                 &84.3$\%$($\pm 4.1\%$)\\
$\Sigma{W(O-C)^2}$  &0.0059\\
\hline
\end{tabular}
\end{center}
\end{table}

\section{Discussions and conclusions}

Although TY Pup was discovered in 1928 as a variable star, it was neglected for photometric study and orbital period investigation. Our photometric solutions indicate that TY Pup is an A-subtype deep-contact binary with a high fill-out factor ($f$ = 84.3$\%$) and a low mass ratio ($q$ = 0.184).
These parameters are close to those derived by Gu et al. (1993) and Deb \& Singh (2011), while the photometric mass ratio differs from the spectroscopic one $q=0.25$ given by Duerbeck \& Rucinski (2007). This may be caused by the fact that the radial-velocity curves of TY Pup given by them were constructed by only a few observations. As we can see in Fig. 4 in the paper of Duerbeck \& Rucinski, the spectroscopic mass ratio mainly depended on one data point. The temperature difference of the two components is very small ($\Delta T$=15 K) with $T_{2}/T_{1} = 1.0022$, this suggests that the system is in thermal contact. In addition, the orbital inclination is about 83.6 deg, this indicates that it is a total eclipsing binary and physical parameters we obtained are reliable. The geometrical structure of TY Pup is plotted in Fig 9. Based on spectroscopic elements determined by Duerbeck \& Rucinski (2007), the absolute parameters of TY Pup are estimated as: $M_{1}=1.650M_{\odot}$, $M_{2}=0.303M_{\odot}$, $a=4.653R_{\odot}$, $R_{1}=2.636R_{\odot}$, $R_{2}=1.373R_{\odot}$, $L_{1}=14.112L_{\odot}$ and $L_{2}=3.862L_{\odot}$.

\begin{figure}
\begin{center}
\includegraphics[angle=0,scale=0.4]{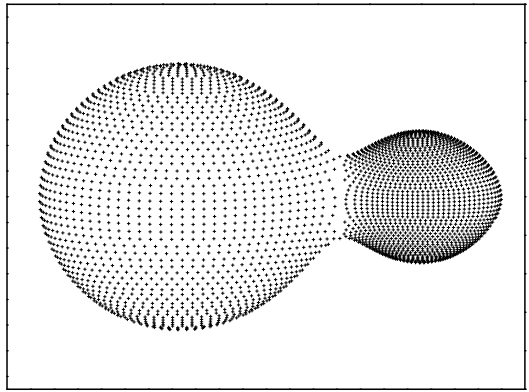}
\caption{Geometrical configuration of TY Pup at phase 0.25}
\end{center}
\end{figure}

\begin{table}
\scriptsize
\caption{Parameters of high fill-out and low mass ratio W UMa binaries.}
\begin{center}
\begin{tabular}{lllccccccccc}\hline\hline
Star &Spec. &Period & $q$ & f &$\mathrm{d}P/\mathrm{d}t$ &Cyclic &$l_3$ & $M_1$ & $M_2$ & $M_3$ &Ref.\\
 & & (days)& & &($\times 10^-{8}$ d/y)& & &($M_{\odot}$)&($M_{\odot}$)&($M_{\odot}$)&\\
\hline
II UMa &F5 &0.8252 &0.172 &86.6$\%$ &+48.8 &no &yes &1.99 &0.34 &1.34 &(1)\\
V2388 Oph &F3V& 0.8023& 0.186& 65.0$\%$ & & & & & & 0.54& (2)\\
MW Pav& F3V&0.7949 &0.222 &60.0$\%$ &+0.06 &no &yes &1.51 &0.33 & & (3)\\
MQ UMa & F7V& 0.4760& 0.195& 82.0$\%$& & yes & yes & 1.33& 0.28& F5V & (4)\\
V409 Hya &F2V &0.4723 &0.216& 60.6$\%$ &+54.1 &no &no &1.50 &0.33 & & (5)\\
V728 Her &F3&0.4713&0.158 &81.0$\%$ &+19.2 &yes&yes &1.80&0.28&0.40&(6)(7)\\
V776 Cas &F2V &0.4404&0.130 &64.6$\%$ & &yes &yes &1.55 &0.20 & 1.04& (8)\\
EF Dra &F9V &0.4240 &0.160 &45.5$\%$ & &yes &yes &1.81 &0.29 &0.75 & (9)\\
TY Pup &F3V &0.8192 &0.184 & 84.3$\%$ & +5.57 &yes &no &1.65 &0.30 &1.12 & this study\\
\hline
\end{tabular}
\end{center}
{\tiny Notes.} \tiny (1) Zhou et al. 2016, (2) Zasche et al. 2014, (3) Alvarez et al. 2015, (4) Zhou et al. 2015, (5) Na et al (2014), (6) Erkan \& Ulas 2016, (7) Yu et al 2016, (8) Zhou et al. 2016a., (9) Pribulla et al. 2001
\end{table}

\begin{table}
\scriptsize
\caption{Parameters of a third body}
\begin{center}
\begin{tabular}{lccc}\hline\hline
 Parameters & Value & Error & Units\\
\hline
$A_3$ &0.0072 &0.0001 &days\\
$P_3$ &3.62 & 0.0000 &yrs\\
$a'_{12} \sin{i'}$ & 1.25 & 0.02 & AU\\
$f(m_3)$ & 0.148 & 0.006 & $M_{\odot}$\\
$e_3$ & 0.0 & assumed & -\\
$M_3$ ($i'=90^{o}$) & 1.117 & 0.020 & $M_{\odot}$\\
$a_3$ ($i'=90^{o}$) & 2.178 & 0.050 & AU\\
\hline
\end{tabular}
\end{center}
\end{table}

The upward parabolic variation in the O-C diagram reveals that the period of TY Pup is increasing continuously at a rate of $\mathrm{d}P/\mathrm{d}t = 5.57(\pm 0.08)\times10^{-8}$ days/year. The period increase can be explained by the mass transfer from the secondary component to the primary one. When material is exchanged between the stars in the system, the center of mass of the system will be shifted and consequently the orbital period of the system will change. If the long-term period increase is due to conservative mass transfer from the less massive component to the more massive one, the mass transfer rate can be determined with the following equation (Tout \& Hall 1991),
\begin{equation}
\frac{\dot{P}}{P} = 3 \frac{\dot{M_2}}{M_2}(1-\frac{M_2}{M_1}).
\end{equation}
The result is $\mathrm{d}M_{2}/\mathrm{d}t = 8.41\times10^{-9}$ $M_{\odot} \textrm{yr}^{-1}$. The timescale of mass transfer can be estimated as $M_2/\dot{M_2} \sim 3.6\times10^7 \textrm{yrs}$ and the time scale of period increase $P/(\mathrm{d}P/\mathrm{d}t) \sim 1.47\times10^7 \textrm{yrs}$ or $\dot{P}/P$ $\sim$ 6.8$\times 10^{-8}$ $\textrm{yr}^{-1}$. If the more massive star ($M_1$) is gaining mass from the less massive star ($M_2$), the mass ratio of the contact binary ($q$) will decrease. The primary will become too massive (Qian 2001b). However, the contact configuration cannot be broken, due to its deep contact configuration with a high fill-out factor, f $>$ 50$\%$ (He et al. 2012). By using the statistical relation between $f$ and $q$ for low-mass-ratio and deep contact binaries derived by Yang \& Qian (2015),
\begin{equation}
f(\%) = 117.6(\pm7.5)-527.6(\pm2.5)\times q + 1164.9(\pm7.4)\times q^2,
\end{equation}
a calculation with the mass ratio $q$=0.184 yields the fill-out factor of TY Pup as $f=59.96\%$. This is smaller than the observed value ($f$ = 84.3$\%$).

The low mass ratio together with the deep contact configuration of TY Pup indicate that it is at the late evolutionary state of contact binaries.
According to Hut (1980), when the W UMa system meets a secular tidal instability, i.e., the orbital angular momentum is less than three times of the spin angular momentum ($J_{orb}$ $<$ $3J_{spin}$), the system will ultimately merge to be a single rapidly rotating star.
A computation with the relation between the mass ratio and the angular momentum ratio ($q-J_{spin}/J_{orb}$) given by Yang \& Qian 2015),
\begin{equation}
J_{spin}/J_{orb} = 0.5104(\pm0.0006)-3.7738(\pm0.0027)\times q + 8.2817(\pm0.0081)\times q^2
\end{equation}
leads to the angular momentum ratio for TY Pup as 0.096. The decrease of $q$ caused by the mass transfer from the less massive component to the more massive one will cause the system finally meet $J_{spin}/J_{orb} > 1/3$. At that time, the binary will be merging and produce a luminous red nova (e.g., Zhu et al. 2016). Some contact binary systems with observational properties similar to TY Pup are listed in Table 6. All of them are F-type deep contact system with mass ratios lower than 0.25 and fill-out factors larger than 50\% (Qian et al. 2005). They may be the progenitor of a single rapid-rotating star and will produce luminous red novae (e.g., Zhu et al. 2016; Sriram et al., 2016, 2017; Liao et al. 2017; Samec et al. 2011, 2018).

The cyclic variation of $O-C$ diagram in Fig 4 can be explained as magnetic activity cycles which normally occur in the late-type stars (e.g. Applegate 1992). However, as discussed by Qian (2001a, 2003), magnetic braking in high fill-out over-contact binaries may be weaker than that in shallow contact binaries. Furthermore, since TY Pup have been found and investigated for decades, no magnetic activity was found from available publications (e.g. Stepien et al. 2001). In addition, it is clear that the variation is periodic, so this variation may be more plausibly interpreted as the light-travel time effect (LTTE) via the presence of a third body. Therefore, we thought that the unseen tertiary may be the reason to cause the cyclic oscillation. To derive the parameters of the third component, we assumed that the tertiary's orbit is circular. The parameters of the third component were determined by using the mass function equation,
\begin{equation}
f(m) = \frac{4 \pi^2}{GP_{3}^2} \times (a'_{12} \sin i')^3=\frac{(M_3 \sin i')^3}{(M_1 + M_2 + M_3)^2},
\end{equation}
where the projected radius of the orbit $a'_{12} \sin{i'}$ = $A_3 \times c$ (when $A_{3}$ is the semi-amplitude of the $O-C$ oscillation, $c$ is the speed of light and ${i'}$ is the inclination of the orbit of the third component). The corresponding results are shown in Table 7.

The lowest mass of the tertiary $M_3$ $\sim$ 1.12 $M_{\odot}$ (at $i'$ = 90). It should be very bright and easily detected with either in photometry and spectroscopy as the same V779 Cas (Zhou et al. 2016a). If the tertiary really exists, it may play an important role for the binary formation and evolution by removing angular momentum from the central binary through Kozai oscillation (Kozai 1962) during the early dynamical interaction or late evolution as discussed by Qian et al.(2007, 2013). However, the photometric solution suggests that the contribution of the third light to the total light of the system is very small. Additionally, no third lights were reported from previous photometric investigations (e.g. Gu et al. 1993; Deb \& Singh 2011). Moreover, no spectroscopic signal of a third component was detected (e.g. Pribulla \& Rucinski 2006; Duerbeck \& Rucinski 2007; D'Angelo et al. 2006; Rucinski et al. 2013) or from APOGEE spectra (e.g. El-Badry et al. 2018). Other investigations by Rucinski et al. (2007) and Zakirov (2010) also showed no third body in the binary. It is possible that the third body might be a compact object. The other possibility is that it may be a close binary containing two very faint component stars. More evidence is needed to prove the existence of a third body in the future.

\bigskip

\vskip 0.3in \noindent
This work is supported by the National Natural Science Foundation of China (No. 11703082). We would like to thank Dr. Wiphu Rujopakarn and NARIT, Thailand for time allocation to use PROMPT-8 for our observations, some observations were obtained by using the 60-cm and 1.0-m telescopes at Yunnan Observatories, China.

\end{document}